\documentclass[aps,twocolumn,showpacs,showkeys]{revtex4}

\begin{document}


\newcommand\beq{\begin{equation}}
\newcommand\eeq{\end{equation}}
\newcommand\beqa{\begin{eqnarray}}
\newcommand\eeqa{\end{eqnarray}}
\newcommand\ket[1]{|#1\rangle}
\newcommand\bra[1]{\langle#1|}
\newcommand\scalar[2]{\langle#1|#2\rangle}

\newcommand\jo[3]{\textbf{#1}, #3 (#2)}


\title{\Large\textbf{Unconditionally Secure Quantum Bit Commitment Using Neutron Double-Slit Interference}}

\author{Chi-Yee Cheung}

\email{cheung@phys.sinica.edu.tw}

\affiliation{Institute of Physics, Academia Sinica\\
             Taipei, Taiwan 11529, Republic of China\\}


\begin{abstract}

Using a neutron double-slit setup, we construct a quantum bit commitment scheme in which time development of quantum states plays an essential role. Our scheme evades the widely accepted no-go theorem by the fact that it is neither possible to stop the time evolution of spreading wave packets, nor is it possible to evolve them backward in time. Moreover, for unstable
particles such as neutrons, one cannot delay detecting their  positions indefinitely, or they would disintegrate spontaneously and escape detection. We find that using non-stationary states instead of stationary ones, unconditionally secure quantum bit commitment is possible.

\end{abstract}



\maketitle


Quantum cryptography is one of the most successful branches
of quantum information science. The well known BB84 quantum
key distribution protocol was proposed early in 1984
\cite{BB84}, and to date it has been implemented in the
real world over distances as large as hundreds of
kilometers \cite{Stucki-2009}. Besides quantum key distribution
there are also many other interesting cryptographic
protocols among which the primitive quantum bit commitment
occupies a special position. The security of quantum bit
commitment (QBC) is an important issue because secure QBC
can be used as an building block for other more complicated
cryptographic protocols \cite{Brassard-96} such as quantum
coin tossing \cite{Blum83} and quantum oblivious transfer
\cite{Bennett-91,Crepeau94,Yao95,Mayers96}. Furthermore it
is known that, at least in classical models, oblivious
transfer can be used to implement secure two-party
computation \cite{Kilian88,Crepeau-95}. So it seems that
the security of QBC also holds the key to this long
standing problem in cryptography as well.

Bit commitment involves two distrusting parties customarily
named Alice and Bob. In the beginning Alice commits to Bob
a secret bit $b\in\{0,1\}$ that is to be revealed at some
later time. In order to ensure Bob that she will keep her
commitment, Alice provides Bob with a piece of evidence
with which he can verify her honesty when she finally
unveils the committed bit. A bit commitment protocol is
secure if it satisfies the following two conditions. (1)
Concealing: Bob gets no information about the value of $b$
before Alice unveils it; (2) Binding: Alice cannot change
$b$ without Bob's knowledge. (In general both conditions
are satisfied only asymptotically as some security
parameter $N$ becomes large.) Furthermore, if the protocol
remains secure even if Alice and Bob had capabilities
limited only by the laws of nature (this is sometimes
referred to as the parties having unlimited computational
power), then it is said to be unconditionally secure.

Consider a simple classical example. Alice writes
down her committed bit $b$ on a piece of paper and locks it in
a box. She gives the box to Bob as evidence of her
commitment but keeps the key to herself. Alice unveils by announcing the value of $b$ and giving the key to Bob for
verification. This protocol seems secure because Alice
cannot change the bit value without access to the box, and
Bob cannot open the box without the key. However, as with
all other classical cryptographic schemes, it is not
unconditionally secure because, e.g., Bob's
ability to open the box by himself cannot be ruled out as a
matter of principle. By introducing quantum mechanics into
bit commitment, one hopes to achieve unconditional security
which is guaranteed by the laws of nature.

In a QBC protocol, Alice and Bob execute a series of quantum and classical operations during the commitment procedure, which finally results in a quantum state $\rho_B^{(b)}$ in Bob's hand. If
 \beq
 \rho_B^{(0)}=\rho_B^{(1)}\label{perfect},
 \eeq
then the protocol is perfectly concealing, and Bob is not
able to extract any information about the value of $b$ from
$\rho_B^{(b)}$. When Alice reveals the value of $b$, she must also provide additional information which, together with $\rho_B^{(b)}$, will allow Bob to check if she is honest.

For some time a QBC protocol \cite{BCJL-93} proposed in
1993 was widely believed to be unconditionally secure, but
it was eventually shown otherwise \cite{Mayers95}.
Moreover, a ``no-go theorem" was put forth in 1997
\cite{LoChau97,Mayers97}, which supposedly proved that  concealing protocols can always be cheated by Alice, consequently
unconditionally secure QBC is ruled out as a matter
of principle. Due to the importance of QBC,
this result, if true, constitutes a major setback for
quantum cryptography.

Before proceeding further, it is instructive to outline the
impossibility proof for the perfectly concealing case \cite{LoChau97,Mayers97}. The proof is based on the assumption that, by quantum entanglement, Alice and Bob can
keep all undisclosed classical information undetermined and stored them at the quantum level. In other words, they can delay all prescribed
measurements without consequences until it is required to
disclose the outcomes. It then follows that at the end of the
commitment phase, there always exists a pure state
$\ket{\Psi^{(b)}_{AB}}$ in the joint Hilbert space
$H_A\otimes H_B$ of Alice and Bob. $\ket{\Psi^{(b)}_{AB}}$
is called a quantum purification of the evidence state
$\rho^{(b)}_B$, such that
 \beq
 {\rm Tr}_A~ \ket{\Psi^{(b)}_{AB}}
 \bra{\Psi^{(b)}_{AB}}
 =\rho_B^{(b)}, \label{reduced}
 \eeq
where the trace is over Alice's share of the state.
Note that whether Bob purifies or not is irrelevant to Alice, and she could simply assume he does.
In general purification requires access to
quantum computers, which is consistent with the assumption
that Alice and Bob have unlimited computational power.

Since the protocol is assumed to be perfectly concealing, therefore Eq. (\ref{perfect}) holds. Then, according to a theorem by Hughston \textit{et al.} \cite{Hughston-93}, the two purifications $\ket{\Psi^{(0)}_{AB}}$ and
$\ket{\Psi^{(1)}_{AB}}$ are related by a local unitary transformation $U_A$ on Alice's side, namely,
 \beq
 \ket{\Psi^{(1)}_{AB}} = U_A \ket{\Psi^{(0)}_{AB}}.
 \label{UA}
 \eeq
Since $U_A$ acts on $H_A$ only, she can implement it without Bob's help. Then it is obvious that Alice can change her commitment at will, and Bob will always conclude that she is honest because his density matrix $\rho_B^{(b)}$ is not affected by $U_A$. So it seems that no QBC protocols could be
simultaneously concealing and binding \cite{LoChau97,Mayers97}.

It has been shown that, if we do not adhere to the standard
cryptographic scenario and allow Alice and Bob to control
two separated sites each, then an unconditionally secure
classical bit commitment protocol can be constructed using
relativistic effects \cite{Kent99}. Nevertheless, in this
paper, we are interested only in non-relativistic quantum
bit commitment in the standard scenario where each party
controls only one site.

Since 1997, there have been many attempts at breaking the
no-go theorem \cite{Yuen07}, however a mathematically
rigorous result is still lacking so far. Conceptually, the
claim that unconditionally secure QBC is ruled out in
principle is rather puzzling to some. First of all, while
the objective of QBC is well defined, the corresponding
procedure is not precisely specified.  That is, there
exists no general mathematical characterization of all
imaginable QBC protocols \cite{Yuen08}. So it is
unlikely that the no-go theorem could have blocked all
possible routes to unconditionally secure QBC. Secondly,
when something in nature is forbidden, there usually exists
a deeper and more general reason or principle behind it.
Thus, for example, electric charge can neither be created
nor destroyed because electromagnetic interactions obey an
exact $U(1)$ gauge symmetry; arbitrary quantum states
cannot be perfectly cloned because information cannot be
transmitted faster than the speed of light; etc. However
unconditionally secure QBC is not known to violate any laws
in physics or information theory, so why should it be strictly ruled out?

 As we saw above, the no-go theorem considers only stationary states which do not have significant time evolution properties. However a complete impossibility proof should take into account the effect of time evolution of quantum states in general. Therefore if a protocol uses non-stationary states to carry quantum information, then it is not covered by the no-go theorem. In the following discussion, instead of stationary states, we shall use particles whose spatial wavefunctions are spreading wave packets (their spins are irrelevant). Moreover they are unstable particles with finite lifetimes.

Consider the following example. Alice and Bob shares a pure maximally entangled state of two particles $\ket{\psi_{AB}(\vec x,t)}$, where $(\vec x,t)$ refer to particle-$A$, and the degrees of freedom of particle-$B$ are not explicitly shown. Particle-$A$'s spatial wavefunction involves spreading wave packets of macroscopic dimension while its spin is irrelevant. It is well known that the wavefunction at $t=t_0$ and $t=t_1$ ($t_1>t_0$) are related by a unitary time evolution operator ${\mathcal U}(t_1,t_0)$, namely,
 \beq
 \ket{\psi_{AB}(\vec x,t_1)}={\mathcal U}(t_1,t_0)
 \ket{\psi_{AB}(\vec x,t_0)},
 \label{evolution}
 \eeq
where
 \beq
 {\mathcal U}(t_1,t_0)=e^{-iH_A(t_1-t_0)},
 \eeq
$H_A=p^2/2m_A$ is the free-particle Hamiltonian operator of particle-$A$, and we have assumed $H_B=0$ for simplicity.
Suppose to commit to $b=0$, Alice is required to determine particle-$A$'s position at $t=t_0$, and for $b=1$ she must do so at time $t=t_1$ ($t_1>t_0$).  Clearly the wavefunctions involved are not stationary, and schemes of this type are therefore not ruled out by the no-go theorem.  We show below that indeed the impossibility proof does not go through.

To proceed, let us follow the cheating strategy suggested by the no-go theorem: For $b=0$, Alice measures the particle position coherently at $t=t_0$ with unitary operation and ancilla (suppose it is possible). The resulting pure state can be written schematically as
 \beq
 \ket{\Psi_{AB}^{(0)}(\vec x,t_0)}=
 M[\,\ket{\Phi}\ket{\psi_{AB}(\vec x,t_0)}],
 \eeq
where $\ket{\Phi}$ is the ancilla state, and $M$ is a local unitary operator which entangles $\ket{\Phi}$ with $\ket{\psi_{AB}(\vec x,t_0)}$. If necessary the outcome of the measurement can be obtained by a projective measurement on the ancilla state. Likewise if she commits to $b=1$,
 \beq
 \ket{\Psi_{AB}^{(1)}(\vec x,t_1)}=
 M[\,\ket{\Phi}\ket{\psi_{AB}(\vec x,t_1)}].
 \eeq
So mathematically, just as the no-go theorem says, there does exist a local unitary transformation $U_A$ connecting $\ket{\Psi_{AB}^{(0)}(\vec x,t_0)}$ and $\ket{\Psi_{AB}^{(1)}(\vec x,t_1)}$, with
 \beq
 U_A=M\,{\mathcal U}(t_1,t_0)\,M^{\dagger}.
 \label{UA}
 \eeq
But it is also clear that Alice cannot naively use this $U_A$ to cheat, because $\ket{\Psi_{AB}^{(0)}(\vec x,t_0)}$ and $\ket{\Psi_{AB}^{(1)}(\vec x,t_1)}$ are not stationary states.  Consequently, by the time Alice unveils at $t=T$, $\ket{\Psi_{AB}^{(0)}(\vec x,t_0)}$ would have evolved to
 \beq
 \ket{\tilde\Psi_{AB}^{(0)}(\vec x,T)}=
 {\mathcal U}(T,t_0)\ket{\Psi_{AB}^{(0)}(\vec x,t_0)},
 \eeq
where we have assumed $H_{\Phi}=0$ for simplicity. Therefore, in order to succeed, Alice must either (i) Artificially evolve $\ket{\tilde\Psi_{AB}^{(0)}(\vec x,T)}$ back to $t=t_0$ when she unveils, or (ii) Freeze $\ket{\Psi_{AB}^{(0)}(\vec x,t_0)}$ at $t=t_0$ when she commits. But such manipulations are clearly not possible even with a quantum computer. Furthermore, even if they \textit{were} possible, if particle-$A$ is unstable with a lifetime $\tau\le T$, then it is not possible for Alice to keep its position undetected until $t=T$, because there is a finite probability ($1-e^{-T/\tau}$) that by then there is no particle to detect.

With this observations, we present below a QBC protocol which is unconditionally secure. We shall formulate it explicitly using neutrons which decay spontaneously into a proton($p$), an electron($e$), and an anti-neutrino($\bar\nu$),
$n\rightarrow p+e+\bar\nu$,
with a mean lifetime of $\tau=885.7$ seconds\cite{PDG-08}.  Nevertheless any unstable particles with a suitable lifetime will do. The required experimental setup is a double-slit apparatus similar to that used to demonstrate the wave nature of neutrons \cite{Zeilinger-88}. So this protocol is realizable with available technologies.
\begin{itemize}
\item[]\hspace{-0.5cm}{Commit:}
\item[1.]{At time $t=0$, Bob generates a neutron with momentum $p$ (its spin is irrelevant) and sends it into a double-slit apparatus. With probability 1/2, he randomly closes one of the slits before the neutron arrives. A detector screen is placed at a distance $L$ from the slits on the exit side of the apparatus. If uninterrupted, the neutron enters the slits at $t=t_0$, and reaches the screen at $t=t_1$. We assume that $t_1<<\tau$, so that the possibility of neutron decay before $t=t_1$ is negligible.}
\item[2.]{To commit to $b=0$, Alice must find out which slit does the neutron go through. For $b=1$, she must detect it on the screen and records its position. At $t=t_1$, Alice must announce whether a neutron has been detected or not (the neutron makes it through the slits with a finite probability $\alpha$).}

\item[3.]{The above procedure is repeated $N$ times. Or, to save time, the whole commitment process can be carried out in parallel with $N$ identical double-slit setups. The latter case is assumed for simplicity, and the commitment phase ends at $t=\tau$.}
\item[]\hspace{-0.5cm}{Unveil:}
\item[1.]{At time $t=T$, Alice unveils the value of $b$ and the corresponding neutron detection data. Specifically, for $b=0$, she must specify the slit from which each of the $\alpha N$ neutrons emerged; for $b=1$, she must reveal the position at which each of these neutrons was detected on the screen.}
\item[2.]{Bob checks if Alice's data are consistent with her commitment.}
\end{itemize}

Having specified the protocol, we proceed to show that it is unconditionally secure.  First of all, this protocol is obviously concealing, since the only information disclosed by Alice during the commitment procedure is if neutrons have been detected. Next we show that it is binding. Let us first consider classical cheating. If Alice had determined the which-slit information at $t=t_0$, then she would not be able to distinguish the single-slit and double-slit events, and consistent reconstruction of the interference pattern on the screen would be impossible. On the other hand, if she had detected the neutrons on the screen at $t=t_1$, the resulting pattern is a superposition of single-slit and double-slit events, and it is impossible to separate the two because they occur randomly. Hence Alice could not commit to one bit value and unveil another. If she did, she could only do so by pure guessing, but then her success probability is only of order $2^{-\alpha N}$.

Next we consider the quantum cheating strategy suggested by the no-go theorem. As shown in Eq. (\ref{UA}), in accordance with the theorem, there is indeed a unitary transformation $U_A$ connecting the wavefunctions corresponding to the $b=0$ and $b=1$ cases, which is essentially just the usual time evolution operator. However, as explained before, this strategy would work only if (i) Alice could freeze the neutron wavefunctions at $t=t_0$, or (ii) she could evolve the wavefunctions from $t=T$ backward to $t=t_0$. Both strategies are obviously impossible. Moreover, even if they \textit{were} possible, there are only  $N(T)=\alpha N e^{-T/\tau}\le\alpha Ne^{-1}$ neutrons left at $t=T$ to be manipulated, as a finite fraction of them have already disintegrated.  Consequently Alice must honestly detect the neutrons at $t=t_0$ or $t=t_1$, otherwise the neutron wavefunctions would evolve (spread and disintegrate) according to the laws of nature, which she cannot stop.
Hence this protocol is binding as well as concealing. This completes the proof that our protocol is unconditionally secure.

From the above discussion, we see that time evolution of non-stationary quantum states plays an essential role in our protocol. In contrast, the time factor is not considered in the no-go theorem \cite{LoChau97,Mayers97}, where the wavefunctions involved are assumed to be stationary. In our protocol, neutron wavefunctions are non-stationary in two senses: (i) The neutron spatial wavefunctions expand in size, and (ii) Neutrons are unstable with a mean lifetime of about 15 minutes \cite{PDG-08}. These features are crucial to the security of our protocol. We emphasize that formally there does exist a unitary transformation $U_A$ connecting the $b=0$ and $b=1$ states, so the no-go theorem is mathematically correct. But Alice cannot use this $U_A$ to cheat because, by the time she unveils, the quantum state has evolved to something beyond her control.

In summary, using wave packets of unstable particles instead of stationary states, we have constructed a new QBC protocol and shown that it is unconditionally secure. Our protocol is basically a neutron double-slit experiment, similar to that used to demonstrate the wave nature of neutrons \cite{Zeilinger-88}. The security of this protocol is due to the facts that it is neither possible to stop the spreading of a wave packet, nor is it possible to evolve it backward in time. Moreover neutrons are unstable, so that detection of their positions cannot be indefinitely delay, or they will disintegrate spontaneously and eventually there will be no neutrons left to be detected. Our result shows that if non-stationary time evolution of quantum states is taken into account, then the widely accepted no-go theorem we can evaded, and unconditionally secure QBC is possible.




\end{document}